\documentclass[figures]{epl} 
\RequirePackage{amsfonts}[1995/01/01]

\def\3{\ss }           
\def\bsigma{\mbox{\boldmath $\sigma$}}
\DeclareMathSymbol{\blacktriangle}  {\mathord}{AMSa}{"48}  

\def\bxi{\mbox{\boldmath $\xi$}}
\def\bsxi{\mbox{\boldmath \tiny $\xi$}}

\title{Consistent particle-based algorithm with \\ a non-ideal 
equation of state} 

\shorttitle{Consistent particle-based algorithm...}
 
\author{T. Ihle\inst{1}, E. T{\"u}zel\inst{2,3} and D.M. Kroll\inst{1,3}}   

\institute{
\inst{1} Department of Physics, North Dakota State University,\\ P.O. Box 5566,
Fargo, ND 58102, USA.\\
\inst{2} School of Physics and Astronomy, 116 Church Street SE,\\ University of Minnesota, Minneapolis, MN 55455, USA. \\
\inst{3}Supercomputing Institute, 
University of Minnesota, \\ 
599 Walter Library, 
117 Pleasant St. SE,
Minneapolis, MN 55455, USA \\
}
\pacs{02.70.Ns}{Molecular dynamics and particle methods}
\pacs{47.11.+j}{Computational methods in fluid dynamics}
\pacs{05.40.-a}{Fluctuation phenomena, random processes, noise, and Brownian motion}

\begin{document}

\maketitle
 
\begin{abstract} 
A thermodynamically consistent particle-based model for fluid dynamics with 
continuous velocities and a non-ideal equation of state is presented. 
Excluded volume interactions are modeled by means of
biased stochastic multiparticle collisions which depend on the local 
velocities and densities. Momentum and energy are exactly conserved locally. 
The equation of state is 
derived and compared to independent measurements of the pressure. Results for 
the kinematic shear viscosity and self-diffusion constants are presented. 
A caging and order/disorder transition is observed at high densities and 
large collision frequency. 
\end{abstract} 
 

\section{Introduction}
The efficient modeling of the long length- and time-scale dynamics of 
complex liquids such as colloidal and polymeric suspensions requires a 
simplified, coarse-grained description of the solvent degrees of freedom.  
A recently introduced particle-based simulation technique 
\cite{male_99}---often called stochastic rotation dynamics (SRD) 
\cite{ihle_01,ihle_03a,ihle_04,tuze_03,kiku_03,pool_05} 
or multi-particle 
collision dynamics \cite{lamu_01,ripo_04}---is a very promising 
algorithm for mesoscale simulations of this type. In additional to 
its numerical advantages, the algorithm  
enables simulations in the microcanonical ensemble, and fully incorporates 
both thermal fluctuations and hydrodynamic interactions. Furthermore, its 
simplicity has made it possible to obtain analytic expressions for the 
transport coefficients which are valid for both large and small mean free 
paths, something that is often very difficult to do for other mesoscale 
particle-based algorithms.This algorithm  
is particularly well suited for studying phenomena with  
Reynolds and Peclet numbers of order one, and it has been used to study 
the behavior of polymers \cite{kiku_02,ripo_04}, colloids 
\cite{male_99,lee_04,padd_04,hecht_05}, 
vesicles in shear flow \cite{nogu_04}, and complex 
fluids \cite{hash_00,saka_02a}. 

The original SRD algorithm models a fluid with an ideal gas equation of state. 
The fluid is therefore very compressible, and the speed of sound, $c_s$,
is low. In order to have negligible compressibility effects, as in real 
liquids, the Mach number has to be kept small, which means that there are 
limits on the flow velocity in the simulation. It is therefore important to 
explore ways to extend the algorithm to model dense fluids. 
Our approach starts from what has been a common theme of most liquid 
theories, namely the separation of intermolecular forces into short- 
and long-range parts, which are then  
treated differently. The short-range component is a strong repulsion when 
molecules are close together; it leads to excluded volume effects which 
cause a decrease in the fluid's compressibility and eventual crystallization 
at low temperatures or high density. The long-range component is a weak 
attraction which can lead to a liquid-gas transition. 
The generic reference system for the short-range repulsive 
component of the force is the hard sphere system.  
In this letter we show how the SRD the algorithm can be modified to model 
excluded volume effects, allowing for a more realistic modeling
of dense gases and liquids. This is done in a thermodynamically consistent 
way by introducing generalized excluded volume interactions between
the fluid particles. The algorithm can be thought of as a coarse-grained 
multi-particle collision generalization of a hard sphere fluid, since, 
just as for hard spheres, there is no internal energy. In order to simplify 
the analysis of the equation of state and the transport coefficients, and 
enhance computational efficiency, the cell structure of the original SRD 
algorithm is retained. This work is a first step towards developing 
consistent particle-based algorithms for modeling, in the microcanonical 
ensemble, more general liquids with additional attractive interactions and a 
liquid-gas phase transition.

\section{Model}

As in the original SRD algorithm, the solvent is modeled by a large number 
$N$ of point-like particles of mass $m$ which move in continuous space with 
a continuous distribution of velocities. The system is coarse-grained 
into $(L/a)^d$ cells of a $d$-dimensional cubic lattice of linear 
dimension $L$ and lattice constant $a$. 
The algorithm consists of individual streaming and collision steps. In 
the free-steaming step, the coordinates, ${\bf r}_i(t)$, of the solvent 
particles at time $t$ are updated according to ${\bf r}_i(t+\tau)=
{\bf r}_i(t) + \tau {\bf v}_i(t)$, where ${\bf v}_i(t)$ is the velocity 
of particle $i$ at time $t$ and $\tau$ is the value of the discretized 
time step. In order to define the collision, we introduce a second grid  
with sides of length $2a$ which (in $d=2$) groups four adjacent cells into one 
``supercell''. 

As discussed in Refs. \cite{ihle_01} and \cite{ihle_03a}, a random shift of 
the particle coordinates before the collision step is required to ensure 
Galilean invariance. All particles are therefore shifted by the {\it same}
random vector with components in the interval $[-a,a]$ before the collision 
step (Because of the supercell structure, this is a larger interval than in 
the conventional SRD algorithm). 
Particles are then shifted back by the same amount 
after the collision. To initiate a collision, pairs of cells in every supercell 
are randomly selected. As shown in Fig. 1, three different choices are 
possible: a) horizontal (with $\bsigma_1=\hat x$), b) vertical ($\bsigma_2
=\hat y$), and c) diagonal collisions (with $\bsigma_3=
(\hat x+\hat y)/\sqrt{2}$ and $\bsigma_4=(\hat x-\hat y)/\sqrt{2}$). Note that 
diagonal collisions are essential to equilibrate the kinetic energies in the 
$x-$ and $y-$directions. 

\begin{figure}
\twofigures[width=2.5in]{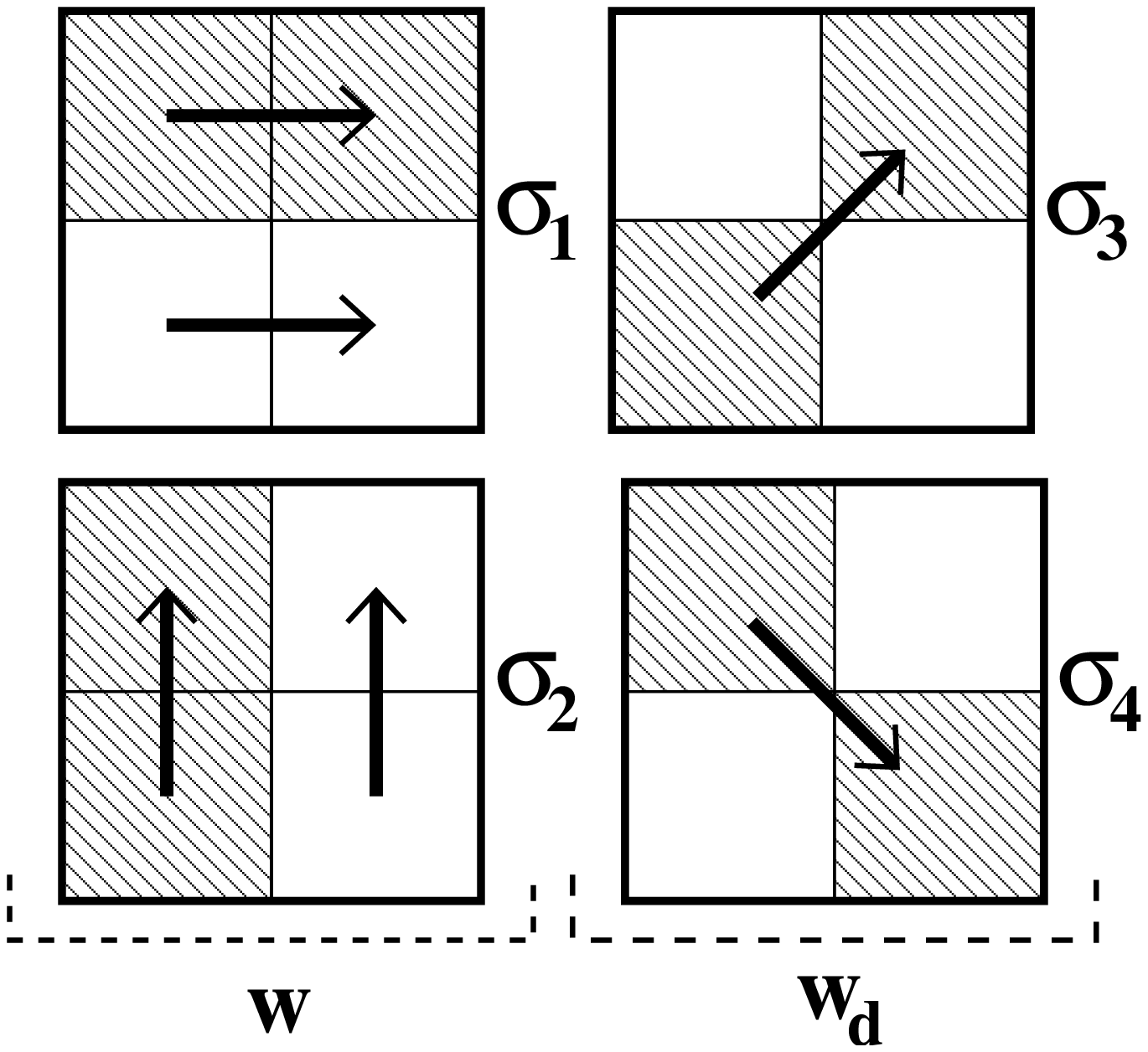}{D_vs_taukT_inset.eps}
\caption{Schematic of collision rules. Momentum is exchanged in four ways, 
a) horizontally along $\bsigma_1$, b) vertically along $\bsigma_2$, 
c) diagonally and d) off-diagonally along $\bsigma_3$ and $\bsigma_4$ 
respectively, according to Eq. (\ref{NONID2}). $w$ and $w_d$ denote the 
probabilities of choosing collisions a), b) and c), d) respectively.}
\label{fig_rules}
\caption{Diffusion coefficient as a function of $\tau$. The data points in 
the inset are data for the shear viscosity measured using Green-Kubo relations, 
as a function of $\tau k_BT$. The solid line shows the analytical result, 
Eq. (\ref{EXACT_DIFF}).
Parameters: $L=64a$, $M=5$, $k_BT=1.0$ and $A=1/60$.}
\label{fig_diffnu}
\end{figure}

In every cell, we define the mean particle velocity,
\begin{equation}
{\bf u}_n={1\over M_n}\,\sum_{i=1}^{M_n}\,{\bf v}_i, 
\end{equation}
where the sum runs over all particles, $M_n$, in the cell with index $n$. 
The projection of the difference of the mean velocities of the selected 
cell-pairs on ${\bf \sigma}_j$, 
$\Delta u={\bf \sigma}_j\cdot ({\bf u}_1-{\bf u}_2)$, 
is then used to determine the probability of collision.
If $\Delta u<0$, no collision will be performed. For positive $\Delta u$, a 
collision will occur with an acceptance probability which depends on 
$\Delta u$ and the number of particles in the two cells, $M_1$ and $M_2$.
This rule mimics a hard-sphere collision on a coarse-grained level: For 
$\Delta u>0$ clouds of particles collide and exchange momenta.  
For reasons discussed in the following, 
we have used the acceptance 
probability  
\begin{equation}
\label{NONID0} 
p_A(M_1,M_2,\Delta u)=\Theta(\Delta u)\,\,{\rm tanh}(\Lambda) 
 \ \ \ \ {\rm with}\ \ \ \ \Lambda = A\,\Delta u\, M_1M_2 ,   
\end{equation}
where $\Theta$ is the unit step function and $A$ is a parameter which 
allows us to tune the equation of state. The hyperbolic tangent function was 
chosen in (\ref{NONID0}) in order to obtain a probability which varies 
smoothly between 0 and 1. 

Once it is decided to perform a collision, an explicit form for the momentum 
transfer between the two cells is needed. The collision should conserve 
the total momentum and kinetic energy of the cell-pairs participating in 
the collision, and in analogy to the hard-sphere liquid, the collision should 
primarily transfer the 
component of the momentum which is parallel to the connecting vector
$\bsigma_j$. In the following, this component will be called the parallel 
or longitudinal momentum. There are many different rules which fullfill these 
conditions. Our goal here is to obtain a large speed of sound. 
We therefore use a collision rule which leads to the maximum transfer of the 
parallel component of the momentum and does not change the transverse momentum.
The rule is quite simple; it exchanges the parallel component of the mean 
velocities of the two cells, which is equivalent to a ``reflection'' of 
the relative velocities,  
\begin{equation}
\label{NONID2}
v_i^{\Vert}(t+\tau)-u^{\Vert}=-(v_i^{\Vert}(t)-u^{\Vert})\,, 
\end{equation}
where  $u^{\Vert}$ is the parallel component of the mean velocity of the 
particles of {\it both} cells. The perpendicular component remains 
unchanged, 
\begin{equation}
v_i^{\perp}(t+\tau)=v_i^{\perp}(t).        
\end{equation}
It is easy to verify that these rules conserve momentum and energy in the 
cell pairs.

Because of $x-y$ symmetry, the probabilities for choosing cell pairs in the 
$x-$ and $y-$ directions (with unit vectors $\bsigma_1$ and $\bsigma_2$ 
in Fig. 1) are equal, and will be denoted by $w$.
The probability for choosing diagonal pairs ($\bsigma_3$ and $\bsigma_4$ in 
Fig. 1) is given by $w_d=1-2w$. $w$ and $w_d$ must be chosen to that the 
hydrodynamic equations are isotropic and do not depend on the orientation 
of the underlying grid. This can be done by considering the temporal 
evolution of the lowest moments of the velocity distribution function. 
It is sufficient to consider the following three moments for a single 
particle $i$, 
\begin{equation}
\Psi_i(t)=
\left( \begin{array}{c}
\langle v_{ix}^2(t) \rangle \\
\langle v_{iy}^2(t) \rangle \\
\langle v_{ix}(t)\,v_{iy}(t)\rangle 
\end{array}
\right) . 
\end{equation}
Assuming molecular chaos, the collision rules can be used to determine the 
eigenvalues of the relaxation matrix, $R$, defined by 
$\Psi_i(t+\tau)=R\,\Psi_i(t)$.   

Because of the conservation of energy, one of the three eigenvalues of $R$ 
is equal to one; the other two are given by $\lambda_1=w_d+2w(2/M-1)$ 
and $\lambda_2=2w+w_d(2/M-1)$, 
where $M$ is the average number of particles per cell. Isotropy requires that 
$\lambda_1=\lambda_2$, a condition that can be fullfiled for arbitrary 
$M$ only if $w_d=1/2$ and $w=1/4$. Simulations show that both the speed 
of sound and the shear viscosity are isotropic for this choice. Note, however, 
that this does not guarantee 
that all properties of the model are isotropic. This becomes apparent 
at high densities or high collision frequency, $1/\tau\gg 1$, where 
inhomogenuous states with cubic or rectangular order can be observed (see 
Fig. 4 and accompanying discussion). 

\section{Transport coefficients} 

The transport coefficients can be determined using the same 
Green-Kubo formalism as was used for the original SRD algorithm \cite{ihle_01,
ihle_03a,ihle_04}. In particular, the kinematic shear viscosity 
is given by 
\begin{equation} 
\label{shear_v}
\nu = \frac{\tau}{Nk_BT}\left.\sum_{n=0}^\infty\right.' 
\langle S_{xy}(0)S_{xy}(n\tau)\rangle, 
\end{equation} 
where  
\begin{equation}
\label{BCOR1}
S_{xy}(n\tau)=\sum_{j=1}^N\,\left(
v_{jx}(n\tau)\Delta\xi_{jy}(n\tau) + \Delta v_{jx}(n\tau)[  
\Delta\xi^s_{jy}(n\tau)-z^s_{jly}([n+1]\tau)/2]\right)   
\end{equation}
is the off-diagonal element of the stress tensor ${\bf S}$.    
${\bxi}_j(t)$ and ${\bxi}^s_j(t)$ are the cell coordinates of particle $j$ 
in the fixed and {\it shifted} frames at time $t$, respectively, 
$\Delta{\bxi}_j(t)={\bxi}_j(t+\tau)-{\bxi}_j(t)$, 
$\Delta{\bxi}^s_j(t)={\bxi}_j(t+\tau)-{\bxi}_j^s(t+\tau)$, and 
$\Delta{\bf v}_j(t) = {\bf v}_j(t+\tau)-{\bf v}_j(t)$. 
${\bf z}^s_{jl}$ indexes pairs of cells which participate 
in a collision event; the second subscript, $l$, is the index of the collision 
vectors $\bsigma_l$ listed in Fig. 1. 
For example, for collisions characterized by $\bsigma_1$, 
$z^s_{j1x}=1$ if $\xi^s_{jx}$ in (\ref{BCOR1}) is one of the two cells on the 
left of a supercell and $z^s_{j1x}=-1$ if $\xi^s_{jx}$ is on the right hand 
side of a supercell; all other components of ${\bf z}^s$ are zero. 
In general, the components of ${\bf z}^s_{jl}$ are either $0$, $1$, or $-1$.  
Using $\{{\bf z}^s_{jl}\}$, 
the collision invariants of the model can be written as   
\begin{equation} 
\label{cons} 
\sum_j \left( {\rm e}^{i{\bf k}\cdot{\bsxi}^s_j(t+\tau)} + 
{\rm e}^{i{\bf k}\cdot({\bsxi}^s_j(t+\tau) + {\bf z}^s_{jl}(t+\tau))} \right)
[a_{\beta,j}(t+\tau) - a_{\beta,j}(t)]=0,
\end{equation}
where $a_{1,j}=1$ for the density, $\{a_{\beta,j}\}=\{v_{\beta-1,j}\}$ are 
components of the particle momentum, and $a_{d+2,j}=v_j^2/2$ is the kinetic 
energy of particle j \cite{ihle_03a}. The analogous collision invariants for 
the standard SRD algorithm are given in Eq. (25) of \cite{ihle_03a}.  
The vectors ${\bf z}^s$ are constructed so that the sum of
the two exponentials in (\ref{cons}) is the same for two particles if and
only if they are in partner cells in a collision with index $l$ (see Fig. 1).

The self-diffusion constant $D$ is given by a sum over the 
velocity-autocorrelation function (see, e.g. Eq. (102) in \cite{ihle_03a}) 
and can be evaluated analytically assuming molecular chaos. 
Due to the excluded volume interactions, density fluctuations are supressed 
in the current algorithm; ignoring these fluctuations, one finds 
\begin{equation}
\label{EXACT_DIFF}
D=k_B T\,\tau\left( {1\over A}\,\sqrt{\pi \over k_B T}\;{M^{-3/2} \over 1+1/(8M) } 
-{1 \over 2} \right)\,,
\end{equation}
which is in good agreement with simulation data, see Fig. 2.

\section{Equation of state}

The collision rules conserve the kinetic energy, so that the internal energy 
of our system should be the same as that of an ideal gas. 
Thermodynamic consistency requires that the 
non-ideal contribution to the pressure is linear in $T$. As will be 
shown, this is possible if the coefficient $A$ in (\ref{NONID0}) is chosen 
small enough (see Fig. 3).  

We use here the mechanical definition of pressure---the average 
longitudinal momentum transfer across a fixed interface per unit time and 
unit surface area---to determine the equation of state. We consider only 
the momentum transfer due to collisions, since that coming from streaming 
constitutes the ideal part of the pressure.

Take an interface that is parallel to the $y-$axis and consider the component 
$p_{xx}$ of the pressure tensor. Only collisions with label $l=1$, $3$, and $4$ 
of the collision vector $\bsigma_l$ in Fig. 1 contribute to the momentum 
transfer in this case. Consider the contribution to the momentum 
transfer across the cell boundary from collisions with $l=1$. 
For fixed number of particles, $M_1$ and $M_2$, in the two cells, 
the thermal average of the momentum transfer, 
$ \Delta G_x $, across the dividing line is 
\begin{equation}
\label{PRESS3}
\langle \Delta G_x \rangle={w\over 2} \int_0^\infty\,p_G(\Delta u)\,
p_A(M_1,M_2,\Delta u) \Delta G_x \,d(\Delta u)\,.  
\end{equation}
The factor $1/2$ comes from the position average of the dividing line, since 
the collision occurs n the shifted cells, and   
the integral is restricted to positive $\Delta u$ because the acceptance rate 
is zero for $\Delta u<0$. $p_G(\Delta u)$ is the probability that 
$u_{1x}-u_{2x}$ for the micro-state of two cells is equal to $\Delta u$.
$w=1/4$ is the probability of selecting this collision. 

\begin{figure}
\twofigures[width=2.5in]{pressure_vs_kTtau.eps}{conf_tau0.001_M5_kT3.125e-5.eps}
\caption{Non-ideal pressure, $P_n$,  as a function of $k_BT/\tau$ averaged over 
$10^5$ time steps. Both $k_B T$ and $\tau$ ranged from $0.005$ to $4$. 
The line represents the theoretical expression, Eq. (\ref{PRESS_MAV}).
For $\tau=0.005$ and $k_B T=1$, $P_n$ is five times larger than $P_{id}$.
Parameters: $L=64a$, $M=5$, $A=1/60$.} 
\label{PRESS_FIG}
\caption{Freezing snapshot after $10^6$ time steps. Parameters: $L_x=L_y=32a$, 
$M=5$, $k_BT=3.125\times10^{-5}$, $\tau=10^{-3}$, $A=1/60$.}
\label{fig_freezing}
\end{figure}

Expanding the acceptance probability, Eq. (\ref{NONID0}), in 
$\Lambda\equiv A\;\Delta u\;M_1\, M_2$ leads to \\
$p_A(M_1,M_2,\Delta u)=\Theta(\Delta u)
(\Lambda-\Lambda^3/3+...)$.
The contributions to the pressure from all terms of this series can be 
calculated, but since the resulting contribution to the pressure from 
a term proportional to $\Lambda^n$ is of order $T^n$, we restrict 
ourselves to the first term. 

The resulting contribution to the pressure, $P(\bsigma_1,M_1,M_2)$, for 
fixed $M_1$ and $M_2$ is the average momentum transfer per unit area and 
unit time, so that using Eqs. (\ref{PRESS3}), 
we have $P(\bsigma_1,M_1,M_2)=w\, A\, k_B T\, M_1 \,M_2/(2a\tau)
 + O(A^3T^2)$. 
A similar calculation can be performed for the contributions from the 
diagonal collisions, which occur with the probability $w_d$. 
Using $w=1/4$ and $w_d=1/2$ and averaging over the number of particles 
per cell (assuming that they are Poisson-distributed and that
the particle number distributions in adjacent cells are not correlated), 
one finds the non-ideal part of the pressure, 
\begin{equation}
\label{PRESS_MAV}
P_n=P_{id}\,\left({1 \over 2 \sqrt{2}}+{1 \over 4} \right) {A\,M \over 2} 
{a \over \tau} + O(A^3T^2) .   
\end{equation}
where 
$P_{id}=k_B T\,M/a^2$ is the ideal gas contribution to the pressure (in $d=2$). 
Note that the same result is obtained if, instead of averaging over $M_1$ and 
$M_2$, we simply set $M_1=M_2=M$, the average number of particles 
per cell. $P_n$ is quadratic in the particle density, $\rho=M/a^2$,   
as one would expect from a virial expansion. The prefactor $A$  
must be chosen small enough that higher order terms in this expansion are 
negligible. We have found that prefactors $A$ leading to acceptance rates of 
about $20\%$ are sufficiently small to guarantee that the pressure is 
linear in $T$ (see Fig. 3). 
In order to measure $P_n$, we have used the fact that the average of 
the diagonal part of the microscopic stress tensor gives the virial expression 
for the pressure 
\begin{equation}
\label{VIRIAL}
P=P_{id}+P_n = \left\langle \sum_j \left\{ 
v_{jx}\Delta \xi_{jx}+\Delta v_{jx}\left[\Delta  \xi^s_{jx}-
z^s_{jlx}/2\right] \right\} \right\rangle .  
\end{equation}
The first term, $\langle v_{x,j}\Delta \xi_{x,j} \rangle=
\langle \tau v_{x,j}^2 \rangle$, gives $P_{id}$, as discussed in Ref. \cite{ihle_03a}. 
The average over the second term vanishes (see Ref. \cite{ihle_03a}), 
while the average of the third term is the non-ideal part of the pressure, 
$P_n$. 
Simulation results for $P_n$ obtained using (\ref{VIRIAL}) are in good 
agreement with the analytical expression, (\ref{PRESS_MAV}) (see 
Fig. \ref{PRESS_FIG}).
In addition, measurements of the density fluctuations, 
$\langle |\rho_k|^2 \rangle$, at small wave vectors ${\bf k}$, 
as well as results for the adiabatic speed of sound obtained from simulations 
of the dynamic structure factor, are both 
in good agreement with the predictions following from Eq. (\ref{PRESS_MAV}). 
These results provide strong evidence for the thermodynamic consistency of the 
model.   

\section{Caging and order/disorder transition}

If the non-ideal part  of the pressure is large compared to the ideal pressure,
ordering effects can be expected. For small $A$, both contributions 
to the pressure are proportional to the temperature, so that just as in a real 
hard-sphere fluid, changing the temperature does not lead to an order/disorder 
transition. On the other hand, the two contributions to the pressure have 
different dependencies on the density and time step, $\tau$. In fact, 
$\tau$ can be interpreted as a parameter describing the efficiency of 
the collisions; lowering $\tau$ results in a higher collision frequency, 
and has a similar effect to making the spheres 
larger in a real hard-sphere system.
We therefore expect caging and ordering effect if either $\rho$ is increased 
or $\tau$ is decreased. This is indeed the case. For $\tau< 0.0016$, $\rho=5$  
and $k_BT= 1.0$, an ordered cubic state is observed. The cubic symmetry of the 
ordered state is clearly an artifact of the cubic cell structure, and it 
would be interesting to see if this could be removed by using an hexagonal 
cell structure or incorporating random rotations of the grid.  
One of the surprising features of this crystalline-like state is, that $x-y$ 
symmetry can be broken.
Furthermore, there is the possibility of having several metastable  
crystalline states corresponding to slightly different 
lattice constants and number of particles per ``cloud''. 
As expected, the lattice constants of these ordered states are 
slightly smaller than the super-cell spacing, $2a$, which sets the range of 
the multi-particle interaction. In this state, the diffusion coefficient 
becomes very small; particles are caged and can barely leave their location.

To understand this behavior, note that without collisions, particle clouds 
will broaden due to streaming; this will happen faster the higher the  
temperature. Due to the grid shift, particles at the perimeter of the 
clouds will more often undergo collisions with neighbor clouds. These 
collisions backscatter the particles, forcing them to fly back 
towards the center of their cloud. There is a correlation between the distance 
from cloud center and rate at which it is backscattered, leading to stable 
cloud formation.
A particle which is left alone between clouds will feel repulsion from 
all clouds and moves around very quickly until it is 
absorbed into a cloud. 

\section{Conclusion}
The model presented in this letter is the first extension of the SRD algorithm 
to model fluids with a non-ideal equation of state. It was shown that the 
model is thermodynamically consistent for the correct choice of acceptance 
probabilities and reproduces the correct isotropic hydrodynamic equations 
at large length scales. Expressions for the equation of state and the self-diffusion constant were derived and 
shown to be in good agreement with numerical results. Simulation results 
for the kinematic viscosity 
were presented, and 
it was shown that there is an ordered state for large densities and 
collision frequencies. 
A detailed analysis of the transport coefficients 
will be presented 
elsewhere. 

\section{Acknowledgement}
Support from the National Science Foundation under Grant No. DMR-0513393 
and ND EPSCoR through NSF grant EPS-0132289 are gratefully acknowledged.
We thank A.J. Wagner for numerous discussions.

\end{document}